\documentclass[preprint]{aastex}
\usepackage{graphicx}
\usepackage{longtable}
\shorttitle{Atlas of MMR in the SS}
\shortauthors{T. Gallardo}

\begin{document}

\begin{center}
\textbf{Evaluating the Signatures of the Mean Motion Resonances in
the Solar System}

Tabar\'e Gallardo

Departamento de Astronom\'ia, Facultad de Ciencias, Igu\'a 4225,
11400 Montevideo, Uruguay

gallardo@fisica.edu.uy
\end{center}

\vspace{4cm}

Submitted to the Journal of Aerospace, Engineering, Sciences and Applications.

\vspace{4cm}

Pages: 15

Tables: 1

Figures: 7

\newpage

\noindent \textbf{Proposed Running head:} Evaluating the Signatures of the Mean Motion Resonances \\

\noindent \textbf{Editorial correspondence to:}\\

\noindent Dr. Tabar\'e Gallardo\\
Departamento de Astronom\'ia,\\
Facultad de Ciencias,\\
Igu\'a 4225,\\
11400 Montevideo,\\
Uruguay\\
Phone: +5982 5258618 ext. 321\\
Fax: +5982 5250580 \\
e-mail: gallardo@fisica.edu.uy

\newpage


{\large \bf Abstract}

\bigskip

The characteristics of the resonant disturbing function for an
asteroid perturbed by a planet in circular orbit are discussed. The
location of the libration centers and their dependence with the
orbital elements of the resonant orbit are analyzed. A proposed
numerical method \citep{ga2006a} for computing the strengths of the
resonances is revised and applied to the region of the main belt of
asteroids showing the relevance of several mean motion resonances
(MMR) with several planets.


%

\newpage


\section{Introduction}

There is an interesting diversity of studies on asteroids in mean
motion resonances (MMR) with Jupiter and on transneptunian objects
in MMR with Neptune. However, not all possible resonances were
analyzed neither all the perturbing planets were considered.

The Solar System is in fact covered by a innumerable quantity of
possible resonances.
 If we do not have a method that
 adequately weighs the strength of each resonance it is laborious to identify which one of the hundreds of
 possible MMRs that
theoretically exist near the semimajor axis of the orbit we are
studying is the one affecting the body's motion.

We know by basic celestial mechanics (see also section 2) that the
strength of a resonance is approximately proportional to the the
mass of the planet and to the eccentricity of the resonant orbit
elevated to the order of the resonance, whenever the eccentricity is
not very high. Consequently, in general we are not motivated to
consider high order resonances neither low mass perturbing planets.

 For zero
inclination orbits and not very high eccentricities it is possible
to compute the widths (somehow related to the strengths) in
semimajor axis of the MMRs with the planets as a function of the
eccentricity \citep{demu83,mal95,nal02} but no simple theoretical
method exists to compute the widths in the case of very eccentric
and non zero inclination orbits.

Recently \citet{ga2006a} presented a numerical method to estimate
the strength of an arbitrary  mean motion resonant orbit with
arbitrary orbital elements assuming a circular orbit for the
perturbing planet. Based on this principle it is possible to compute
the strength of the resonances with all the planets form Mercury to
Neptune for all ranges of semimajor axis, from the Sun up to the
limits of the Solar System. That tool allowed the author to identify
candidates to be in exotic resonances like 6:5 and 1:2 with Venus
and 1:2 and 2:5 with Earth. Very recently a numerous population of
asteroids in the resonance 1:2 with Mars  was also identified
\citep{ga07}.

We resume here the principles of the method, we analyze some
consequences  and we apply it to the main belt of asteroids putting
in evidence the signatures of some MMRs.

\section{The Disturbing Function}

Given a planet of mass $m_P$  and radius vector $\mathbf{r}_P$ in an
heliocentric frame and a small body at $\mathbf{r}$ the equation of
motion is given by
\begin{equation}\label{ecua}
\mathbf{\ddot{r}}+k^2M_{\odot}\frac{\mathbf{r}}{r^3}=\nabla
\mathbb{R}
\end{equation}
where $\mathbb{R}$ is the disturbing function:
\begin{equation}\label{defr}
\mathbb{R}=k^2m_P\Bigl(\frac{1}{\mid \mathbf{r}_P-\mathbf{r}\mid} -
\frac{\mathbf{r}\cdot\mathbf{r}_P}{r^3_P} \Bigr)
\end{equation}

In order to construct an analytical theory for the dynamics of a
small body we need an expression for $\mathbb{R}$. Since Laplace's
times classical expressions were constructed as series expansions
around $e=0$ and $i=0^o$ being one of the most recent versions  the
one of
 \citet{em00}. These expansions have convergence problems for high
 $e$  \citep{fm94} and it is necessary to take into account several terms for
 properly account for high inclination orbits. Planar orbits with high
 eccentricities are best handled by other expansions like Beauge's
\citep{be96}.
 For the general problem a local expansion can be constructed in order
 to study the motion around a small region of the phase space
 \citep{fms89,roigal98}. These \textit{asymmetric} expansions are valid around a small
 region near the center of the expansion but they can be applied to
 very high eccentricity orbits allowing the calculation of the
 precise positions of the
 libration centers and the periods of the small amplitude librations.

 Considering a system
composed by the Sun, the planet and a small body with orbital
elements $(a,e,i,\varpi,\Omega)$ the classical expression of the
expansion for $\mathbb{R}$ is a series of terms of the form:

\begin{equation}\label{df}
\mathbb{R} = \sum C \cos(\varphi)
\end{equation}
being $C$ a function of the form
\begin{equation}\label{C}
C = A(\alpha) e_P^{k1} e^{k2} s_P^{k3} s^{k4}
\end{equation}
with $s=\sin(i/2)$, $A(\alpha)$ being a function of
$\alpha=(a/a_P)^{\pm 1}$ with positive exponent for resonances
interior to planet's orbit and negative for exterior ones. The angle
$\varphi$ is defined as
\begin{equation}\label{angulo}
\varphi = j_1\lambda_P + j_2\lambda + j_3\varpi_P  + j_4\varpi +
j_5\Omega_P + j_6\Omega
\end{equation}
where subscript $P$ denotes the planet.  The $j_i$ are integers
verifying $\sum j_i =0$ with $j_5+j_6$ being always even (D'Alembert
rules) and $k_i\geq \mid j_i \mid$. If $\varphi$ is a quick varying
angle the total effect of the corresponding term of $\mathbb{R}$
will vanish in the long term evolution. Also if $\varphi$ is a slow
varying angle but $C$ is vanishingly small the term again will not
have a dynamical effect in the motion of the particle. Then if we
are interested in a correct description of the long term dynamical
evolution of the particle we must take into account all the slow
varying terms with non negligible coefficients $C$; these are the
resonant and secular terms. As $e_P,e,s_P,s$ are less than 1 we say
that the corresponding term is of order $(k_3+k_4+k_5+k_6)$.
 It is possible to simplify the
analysis taking a circular orbit for the planet with zero
inclination, that means $e_P=s_P=0$. With this approximation the
number of terms involved in (\ref{df}) drop considerably. It is also
possible to take into account in $\mathbb{R}$ the terms due to all
planets.

Under these hypothesis and taking into account D'Alembert rules a
$q$-order resonance $|p+q|:|p|$ with $p$ and $q$ integers occurs
when the general critical angle
\begin{equation}\label{sigmar3bp}
\sigma_j = (p+q)\lambda_P - p\lambda  - (q-2j)\varpi - 2j\Omega =
\sigma + 2j\omega
\end{equation}
librates or have a slow time evolution, where
\begin{equation}\label{sigma0}
\sigma=(p+q)\lambda_P - p\lambda -q\varpi
\end{equation}
is the principal critical angle and $j$ is an integer positive or
negative.
 Due to the very slow time evolution of the angles $(\varpi,\Omega)$ the librations of $\sigma_j$ occur
approximately for
\begin{equation}\label{n}
\frac{n}{n_P} \simeq \frac{p+q}{p}
\end{equation}
where the $n$'s are the mean motions. Then, the formula
\begin{equation}\label{alfa}
\frac{a}{a_P} \simeq (1+m_P)^{-1/3} \Bigl(\frac{p}{p+q}\Bigr)^{2/3}
\end{equation}
defines the location of the resonances with planet $P$ in semimajor
axis. At very low eccentricities the time variation of $\varpi$
cannot be ignored and the location of the resonances are shifted
(the \textit{law of structure}) respect to Eq. (\ref{alfa}). The
integer $p$ is  the degree of the resonance with $p<0$ for exterior
resonances and $p>0$ for interior resonances. With this notation the
trojans (or co-orbitals) correspond to $p=-1$ and $q=0$. The
resonant motion is generated when there is a strong dependence of
$\mathbb{R}$ on $\sigma$ which must be librating or in slow
time-evolution. In this case $\mathbb{R}(\sigma)$ dominate the time
evolution of the small body's orbit.

\section{Properties of the Resonant Disturbing Function}

The limitations imposed by the problems of convergence of the
analytical developments motivate the authors to explore the
disturbing function numerically. In analogy with \citet{schu68}, in
order to explore numerically the function $\mathbb{R}$ for a
specific resonance defined by a semimajor axis given by Eq.
(\ref{alfa}) we can eliminate all short period terms on $\mathbb{R}$
computing the mean disturbing function
\begin{equation}\label{rmean}
R(\sigma)=\frac{1}{2\pi|p|}\int_{0}^{2\pi|p|}\mathbb{R}(\lambda_P,\lambda(\lambda_P,\sigma))d\lambda_P
\end{equation}
for a given set of fixed values of $(e,i,\varpi,\Omega,\sigma)$
where we have expressed $\lambda=\lambda(\lambda_P,\sigma)$ from Eq.
(\ref{sigma0}) with $\sigma$ as a fixed parameter and where
$\mathbb{R}(\lambda_P,\lambda)$ is evaluated numerically from Eq.
(\ref{defr}) where $\mathbf{r}_P$ and $\mathbf{r}$ were expressed as
functions of the orbital elements and mean longitudes $\lambda_P$
and $\lambda$. We repeat for a series of values of $\sigma$ between
$(0^{\circ},360^{\circ})$ obtaining a numerical representation of
the resonant disturbing function $R(\sigma)$.

The equations of the resonant motion show that  the time evolution
of the semimajor axis, $da/dt$, is proportional to $\partial
R/\partial \sigma$ then the shape of $R(\sigma)$ is crucial because
it defines the location of stable and unstable equilibrium points.
For specific values of  $(e,i,\omega)$  the minima of $R(\sigma)$
define the stable equilibrium points also known as \emph{libration
centers} around which there exist the \textit{librations}, that
means oscillations of the critical angle $\sigma$. The unstable
equilibrium points are defined by the maxima.

At low $(e,i)$ the function $R(\sigma)$ calculated from Eq.
(\ref{rmean}) is very close to a sinusoid with amplitude
proportional to $e^q$ as one can expect from the classical series
expansions. At higher eccentricities the orbit approaches to the
planet's orbit, $R(\sigma)$ start to depart from the sinusoid and
classical series expansions start to fail with some exceptions like
Beauge's expansion. For eccentricities greater than the
\textit{collision eccentricity} $e_c$:
\begin{equation}\label{elim}
 e_c \simeq \left| 1 - \Bigl(\frac{p+q}{p}\Bigr)^{2/3} \right|
\end{equation}
the orbit can intersect the planet's orbit and for low inclination
orbits two peaks start to appear around the point where $R(\sigma)$
has its maximum for $e<e_c$. If the two peaks  can be distinguished
then a stable equilibrium point appears between them. For high
inclination orbits the intersection between orbits is less probable
and soft maxima can appear instead of the peaks.
For small
inclination orbits the libration centers are almost independent of
$(i,\omega)$ and can be classified in only three different classes
as showed at Table \ref{tabla}.

Resonances of the type 1:n including trojans (that means 1:1)
exhibit a similar general  behavior. For $e<e_a$, where $e_a$ is
certain value verifying  $e_a<e_c$, there is  a libration center at
$\sigma=180^{\circ}$ and for $e>e_a$ there appear the
\textit{asymmetric} libration centers \citep{be94} with locations
depending not only on $e$ but also on $(i,\omega)$. For trojans we
have $e_a=e_c=0$ so the low eccentricity librations around
$\sigma=180^{\circ}$ do not exist.
 For $e>e_a$ \textit{horseshoe} (HS) trajectories wrapping the
asymmetric librations can exist. These HS trajectories are of the
same nature of the horseshoe trajectories
 in the case of trojans and are only possible for this kind of resonances. In HS trajectories $\sigma$ is
 oscillating with high amplitude around $180^{\circ}$.
For $e>e_c$ at low inclinations they appear two peaks (unstable
equilibrium points) and a stable libration center at
$\sigma=0^{\circ}$. This last equilibrium point is associated with
the known \textit{quasi-satellites} (QS) of the 1:1 resonances
\citep{wie2000}.

For low inclination orbits all odd order interior resonances show
librations around $\sigma=0^{\circ}$, and for $e>e_c$ it appears
another libration point at $\sigma=180^{\circ}$. Conversely, all
interior resonances of order even and all exterior resonances except
resonances of type 1:n show librations around $\sigma=180^{\circ}$,
and for $e>e_c$ it appears another libration point at
$\sigma=0^{\circ}$.

\begin{figure}
 \includegraphics[width=14cm,angle=0]{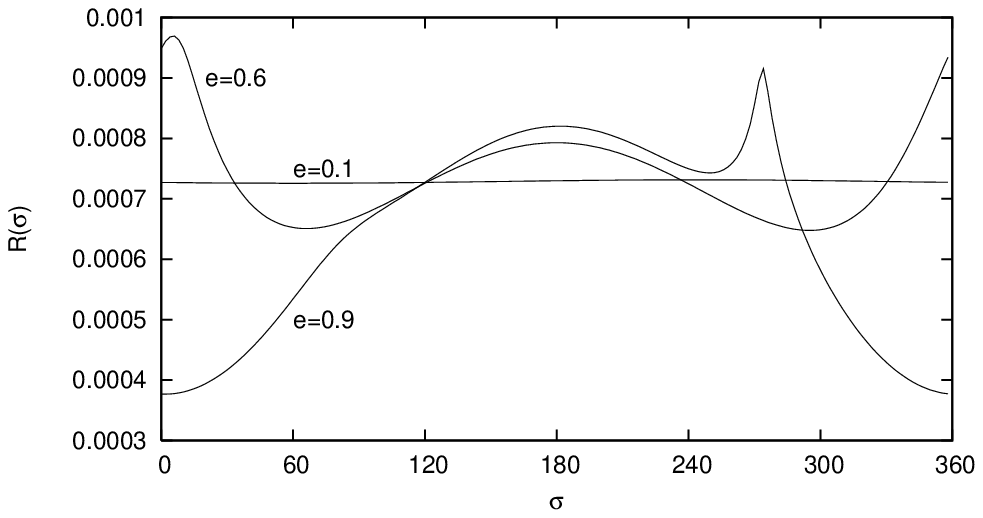}
  \caption{Resonant disturbing function for the resonance 1:4 with Neptune calculated from Eq. (\ref{rmean}) for three
  different values of the eccentricity. In all cases $i=30^o$ and $\omega=60^o$. The minima define the location of the stable
  equilibrium points. Shallow minima at $e=0.1$ implies low stability.}
\label{re}
\end{figure}

\begin{figure}
 \includegraphics[width=14cm,angle=0]{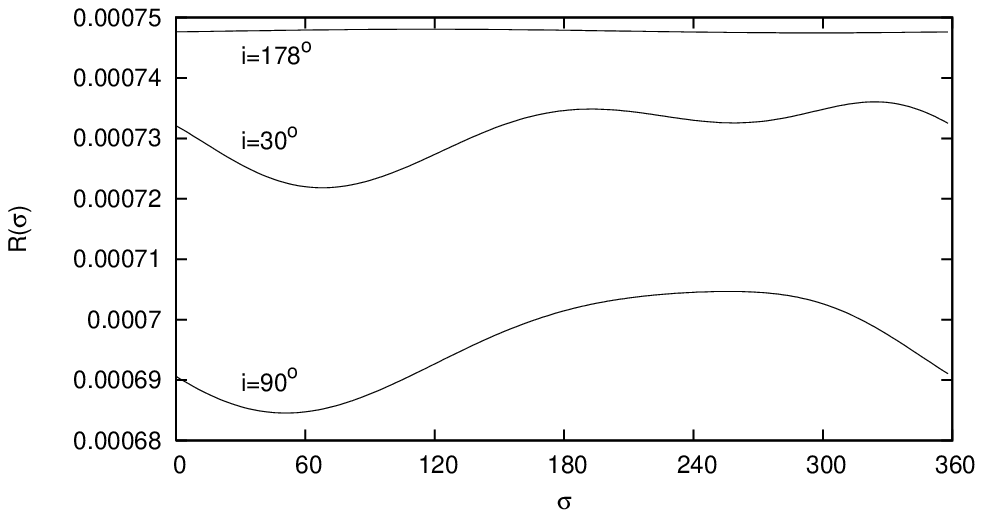}
  \caption{Resonant disturbing function for the resonance 1:4 with Neptune  calculated from Eq. (\ref{rmean}) for three
  different values of the inclination. In all cases $e=0.3$ and $\omega=60^o$. The minima define the location of the stable
  equilibrium points. Shallow minima at $i=178^o$ implies low stability.}
\label{ri}
\end{figure}

\begin{figure}
 \includegraphics[width=14cm,angle=0]{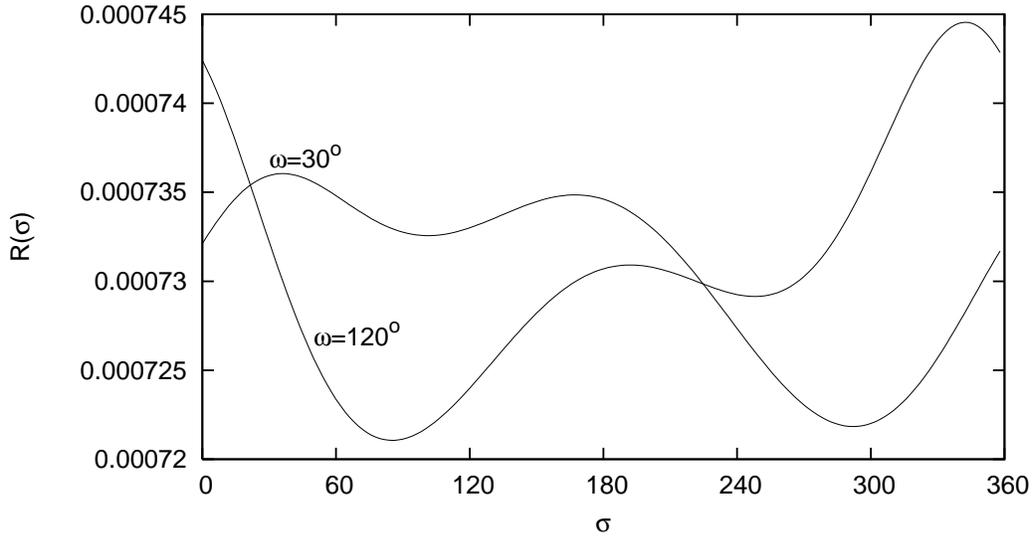}
  \caption{Resonant disturbing function for the resonance 1:4 with Neptune calculated from Eq. (\ref{rmean})
  for two
  different values of the argument of the perihelion. In all cases $e=0.3$ and $i=30^o$. The minima define the location of the stable
  equilibrium points.}
\label{rw}
\end{figure}

For high inclination orbits the geometry of the encounters is
strongly modified affecting the peaks of $R(\sigma)$ and its shape
becomes completely different to the low inclination case becoming
the libration centers strongly dependent on $\omega$. Figures
\ref{re}, \ref{ri} and \ref{rw} illustrate the dependence of
$R(\sigma)$ and its libration centers with $(e,i,\omega)$ for the
case of the exterior resonance 1:4 with Neptune. In all figures we
have taken such units that $k^2m_{Jup}=1$.

\section{The Strength of the Resonances }

\begin{figure}
 \includegraphics[width=14cm,angle=0]{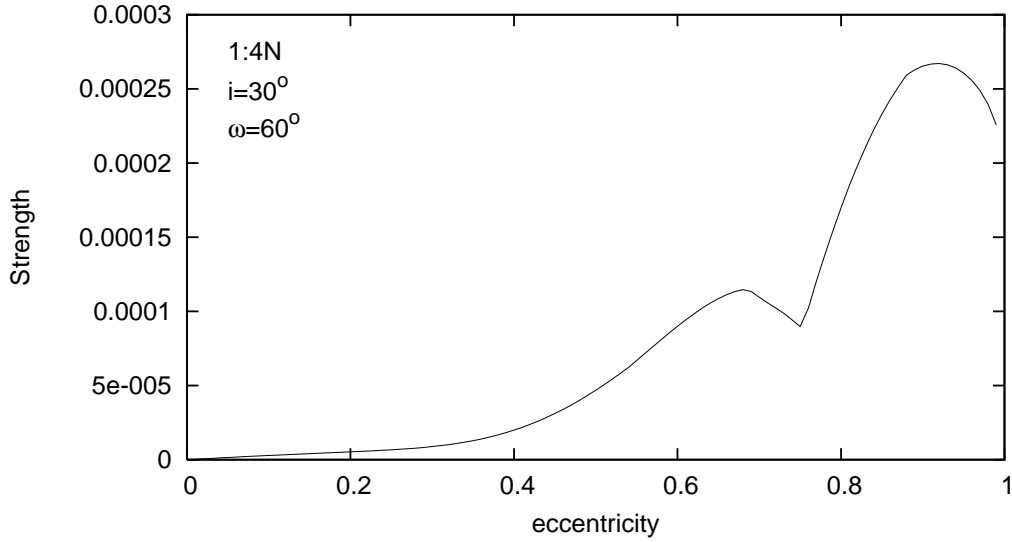}
  \caption{Dependence of the strength $SR(e)$ with the eccentricity for the resonance 1:4 with Neptune  calculated from Eq. (\ref{dr}).
  At low eccentricity regime $SR\propto e^q$ with $q=3$ in this resonance.} \label{fuerzae}
\end{figure}

\begin{figure}
 \includegraphics[width=14cm,angle=0]{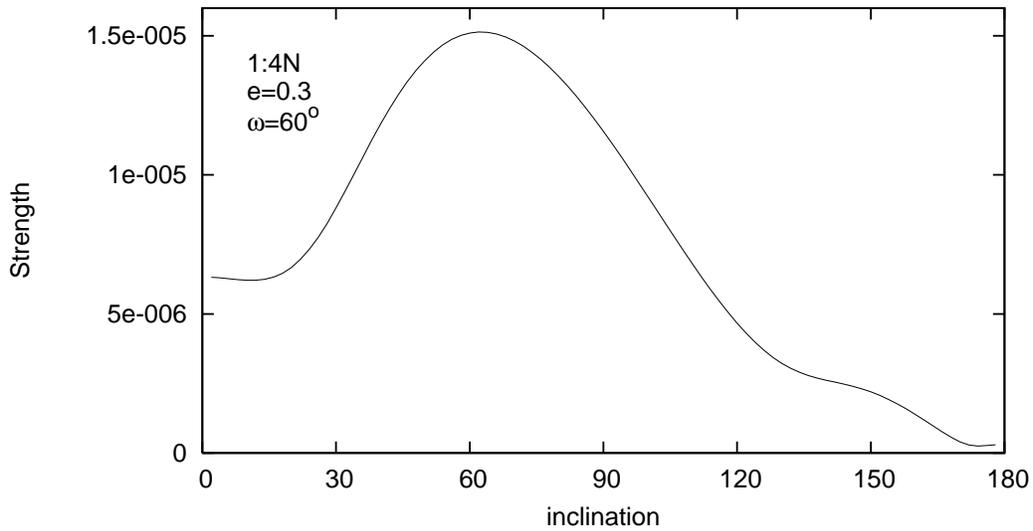}
  \caption{Dependence of the strength $SR(i)$ with the inclination for the resonance 1:4 with Neptune calculated from Eq. (\ref{dr}).
  The strength is in general greater  for high inclination (but direct) orbits.} \label{fuerzai}
\end{figure}

\begin{figure}
 \includegraphics[width=14cm,angle=0]{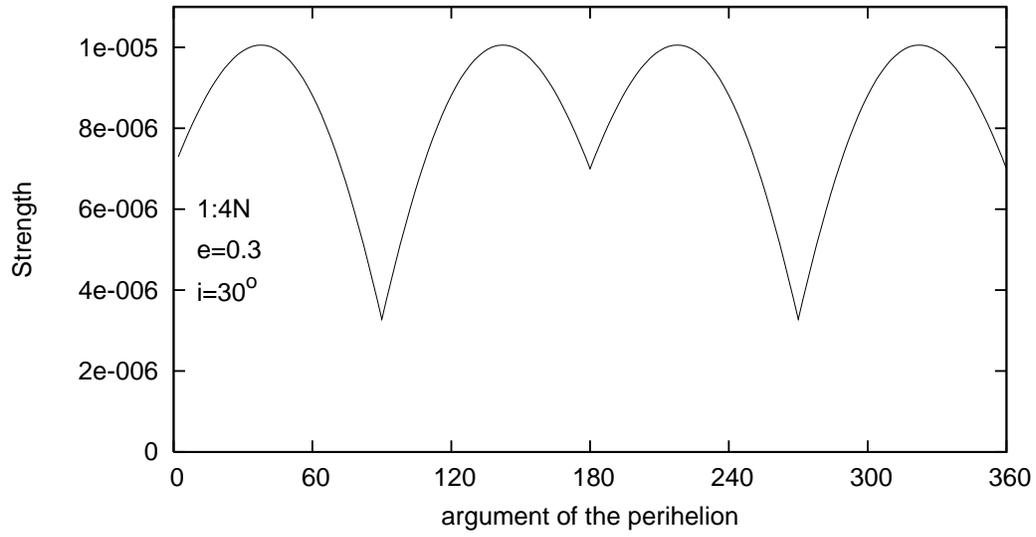}
  \caption{Dependence of the strength $SR(\omega)$ with the argument of the perihelion for the resonance 1:4 with Neptune calculated from Eq. (\ref{dr}).
  The argument of the perihelion affects the location of the equilibrium points but its influence in the
  strength is the less important.} \label{fuerzaw}
\end{figure}

For a given resonant orbit defined by parameters
$(e,i,\varpi,\Omega)$ with a given planet the disturbing function
$R(\sigma)$ is determined. \citet{ga2006a} defined the strength
function $SR$ as:
\begin{equation}\label{dr}
SR(e,i,\omega)=<R>-R_{min}
\end{equation}
being $<R>$ the mean value of $R(\sigma)$ with respect to $\sigma$
and $R_{min}$ the minimum value of $R(\sigma)$. This definition is
in agreement with the coefficients of the resonant terms of the
expansion of the disturbing function for low $(e,i)$ orbits because
for this case $R(\sigma)$ is a sinusoid with an amplitude given by
$<R>-R_{min}$. Then, for low eccentricity and low inclination orbits
$SR$ should follow the function $e^q$. For high $(e,i)$ resonant
orbits the strength cannot be calculated by analytical developments
and departures form the low eccentricity regime is the rule. Figures
\ref{fuerzae}, \ref{fuerzai} and \ref{fuerzaw} illustrate the
dependence of $SR$  with $(e,i,\omega)$ for the case of the exterior
resonance 1:4 with Neptune.

When $SR\sim 0$ we have $\partial R/\partial\sigma \sim 0$ for all
values of $\sigma$ and then $da/dt$ will  not be dominated by
resonant terms
but by other terms that will generate some time
evolution of the semimajor axis and consequently the resonance will
be broken, then the resonance will not be dynamically significant or
strong. On the contrary, a high value of $SR$ implies a strong
dependence of $R$ on $\sigma$ and the resonant disturbing function
$R(\sigma)$ will dominate the motion forcing the semimajor axis to
evolve oscillating around the stable equilibrium points or to evolve
escaping from the unstable equilibrium points.

 \citet{ga2006a}
 analyzed the shape of $SR(e,i,\omega)$ for several resonances and
found that all them can be roughly classified in two classes ($q\leq
1$ and $q\geq 2$) according to the response of $SR$ to the variation
of the inclination which is more important for resonances of order 2
or greater. It is possible to understand why the inclination is an
important factor for resonances of order 2 or greater. Analytical
developments of $R(\sigma)$ in powers of $(e,i)$ show that for a
$q$-order resonance the lowest order resonant terms are of order $q$
in $(e,i)$ \citep{md99}. In particular for trojans \citep{mo99} and
first order resonances the lowest order terms are independent of
$i$. But, for resonances of order $q\geq 2$ the lowest order
resonant terms have a dependence with $i$ that make some
contribution to $R(\sigma)$ for high inclination orbits
\citep{ga2006b}.  In consequence is natural that for resonances of
order 2 or greater  the resonances are stronger for high inclination
orbits because the resonant terms depending on $i$ will show up. On
the contrary we cannot expect such behavior for resonances of order
1 or 0 because the resonant terms depending on $i$ have lower
relevance.

\section{Applications: Identification of the Resonant Signatures}

Following the numerical procedure we have described, \citet{ga2006a}
presented an atlas of MMRs for the Solar System and identified
several objects in unusual resonances with the terrestrial and with
the jovian planets. In particular it was found there exist asteroids
evolving in the exterior resonances 1:2 and 2:5 with the Earth.

\begin{figure}
 \includegraphics[width=16cm,angle=0]{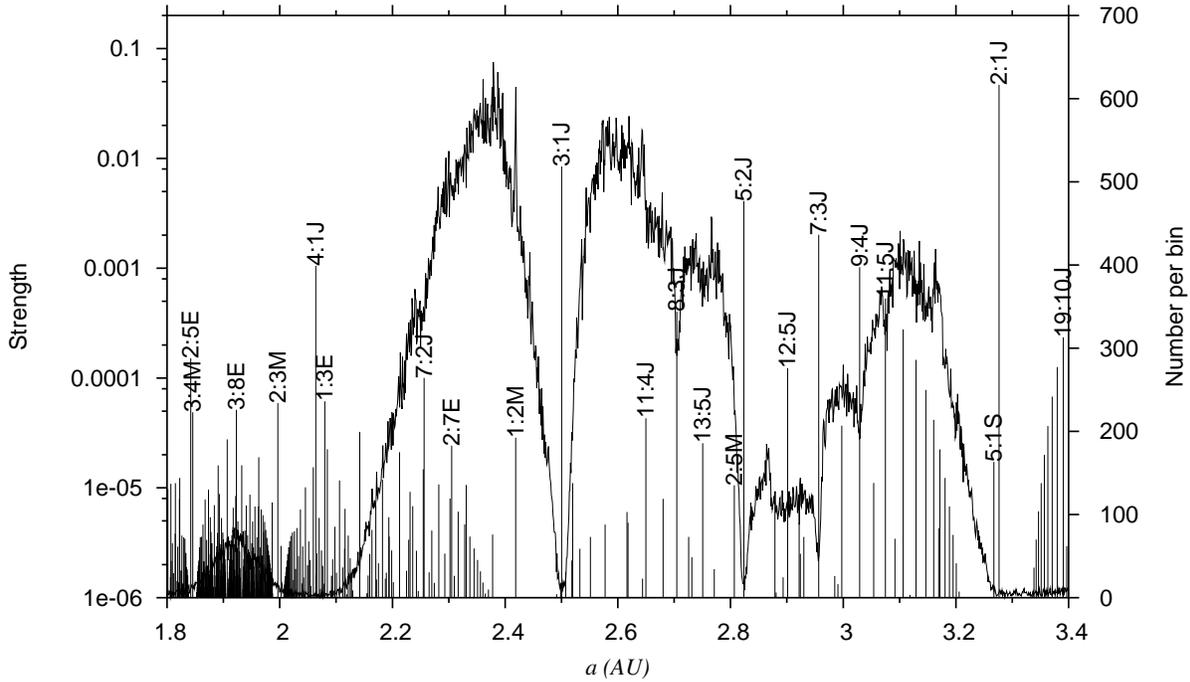}
  \caption{Atlas of the strongest MMRs with all the planets in the region of the main belt of asteroids where strengths were
  calculated from Eq. (\ref{dr}) assuming $e=0.3$, $i=10^o$ and $\omega=60^o$.
  Superimposed is showed an histogram of semimajor axes constructed with bins of 0.001 AU.
The peak at resonance 1:2 with Mars is clearly distinguished.
  This figure was recomposed from \citet{ga07}.} \label{belt}
\end{figure}

We present here an atlas of resonances for the main belt of
asteroids reworked from \citet{ga07} and we compare it with the
distribution of known asteroids (Fig. \ref{belt}). The atlas was
calculated from Eq. (\ref{dr}) considering $e=0.3$, $i=10^o$ and
$\omega=60^o$. The histogram of asteroids was elaborated using the
osculating orbital elements of $\sim 370000$ asteroids taken from
ASTORB database (ftp://ftp.lowell.edu/pub/elgb/astorb.html) and
grouped in bins of 0.001 AU. It is possible to identify the
signatures of some well known resonances with Jupiter. In general, a
secular evolution inside these resonances drives the eccentricity to
values that a collision with the Sun or a close encounter with Mars
or Earth remove the asteroid from the resonance so a gap is
generated.

The main belt of asteroids is limited at its extremes by the
resonances 4:1 (more precisely a secular resonance is responsible
for this border) and 2:1 with Jupiter. The depletion effects due to
resonances 3:1, 8:3, 5:2, 7:3, 9:4 and also 11:5 with Jupiter are
evident. But we can also focus in some signatures not so evident. It
is possible to identify some excess of asteroids in the location of
certain resonances. In particular, in the histogram there is an
excess of around 30\% of asteroids  at $a\sim 2.419$ AU, exactly
where the resonance 1:2 with Mars is located. Note also that the
resonance is isolated and consequently not perturbed by others, that
means, it should dominate in that region. That population inside the
resonance was confirmed via numerical integrations by \citet{ga07}
constituting around a thousand of known asteroids evolving in the
resonance. This is the first numerous population that we have
knowledge captured in a MMR with a terrestrial planet.

Figure \ref{belt} is also showing that resonances 2:5E already
studied by \citet{ga2006a} and 3:4M, 3:8E, 2:3M and 1:3E are strong
enough to be considered possible reservoirs of asteroids, at least
temporarily. This last one probably is strongly perturbed by 4:1J
but 2:3M is relatively isolated and 3:8E is at the middle of a
considerable population of asteroids, in consequence they should be
populated.

As a closing comment, it is evident that analytical theories plus
numerical procedures give us at present a quite complete
understanding of the MMRs and several features of the distribution
of asteroids' populations in the Solar System can be understood in
this context.

\textbf{Acknowledgments}

We acknowledge to the organizers of the XIII Col\'oquio Brasileiro
de Din\^{a}mica Orbital where some of the results included here were
presented. We specially acknowledge A. F. Bertachini de Almeida
Prado for the invitation to contribute in this issue. This work was
developed in the framework of the "Proyecto CSIC I+D, Dinamica
Secular de Sistemas Planetarios y Cuerpos Menores".

\bibliographystyle{amsplain}

\newpage

\begin{table}
\begin{tabular}{|c||c|c|c|}
  \hline
  Resonance Type &   $\sigma_0$ & new $\sigma_0$ for $e>e_c$ \\
  \hline
  \hline
  exterior 1:n  & $180^{o}$ or asymmetric & $0^{o}$ \\
  \hline
  exterior others    &    &    \\
  and &  $180^{o}$  & $0^{o}$  \\
   interior $q$ even  &    &   \\
  \hline
  interior $q$ odd &  $0^{o}$  & $180^{o}$  \\
  \hline
\end{tabular}
  \caption{Stable equilibrium points for low inclination resonant orbits. Resonances of type 1:n except trojans have
  an equilibrium point at $\sigma=180^o$ for low eccentricity orbits. For higher eccentricities this
  point bifurcates in the two asymmetric points. For eccentricities greater than the collision eccentricity, $e_c$, it appear another new equilibrium point in
  all resonances. For high inclination orbits this scheme is strongly modified and the argument of the perihelion
  becomes relevant for the location of the equilibrium points (see Fig. \ref{rw}).}\label{tabla}
\end{table}

\end{document}